\documentclass[draft,jgrga]{agutex}

\usepackage[pdftex]{graphicx}

\setkeys{Gin}{draft=false}
\authorrunninghead{HOLAPPA ET AL.}

\titlerunninghead{INDEPENDENT COMPONENT ANALYSIS}

\begin{document}

%
%

\title{A new method to estimate annual solar wind parameters and contributions of different solar wind structures to geomagnetic activity}

%

%
%


\authors{L. Holappa\altaffilmark{1}, K. Mursula\altaffilmark{1}, T. Asikainen\altaffilmark{1}} 
\altaffiltext{1}{ReSoLVE Centre of Excellence, Department of Physics, FIN-90014 University of Oulu, Finland}







%
%


\begin{abstract}

In this paper, we study two sets of local geomagnetic indices from 26 stations using the principal component (PC) and the independent component (IC) analysis methods. 
We demonstrate that the annually averaged indices can be accurately represented as linear combinations of two first components with weights systematically depending on latitude.
We show that the annual contributions of coronal mass ejections (CMEs) and high speed streams (HSSs) to geomagnetic activity are highly correlated with the first and second IC. 
The first and second ICs are also found to be very highly correlated with the strength of the interplanetary magnetic field (IMF) and the solar wind speed, respectively, because solar wind speed is the most important parameter driving geomagnetic activity during HSSs while IMF strength dominates during CMEs.
These results help in better understanding the long-term driving of geomagnetic activity and in gaining information about the long-term evolution of solar wind parameters and the different solar wind structures.

\end{abstract}

%
%

%

\begin{article}

%
%

\section{Introduction}

Geomagnetic activity is produced in the interaction between the solar wind and the Earth's magnetic field. 
It has been studied systematically since the late 19th century using different geomagnetic indices. 
Most common geomagnetic indices are global indices such as aa, Kp/Ap, Dst and AE, which are constructed from local indices, e.g., as weighted or normalized averages.
For example, the Kp index is calculated from local K indices of 13 magnetic observatories located at midlatitudes and subauroral latitudes. 
Local geomagnetic indices are mainly used to derive global indices but the differences between local indices are rarely studied. 
This is surprising since there are over 200 magnetic observatories around the world continuously producing magnetic measurements, but the state of the Earth's magnetic field is often described by just one globally averaged number.

It has been known for a long time that global geomagnetic activity (measured, e.g., by the aa index) exhibits a dual peak structure during the solar cycle \citep{Chapman_Bartels_1940, Newton_48}, the first peak during the solar maximum dominated by transient activity and the second peak during the declining phase related to recurrent activity.  
Later it became clear that the first peak is mainly produced by coronal mass ejections (CMEs) and the second peak mainly by high speed streams (HSSs) \citep{Simon_Legrand_86, Gosling_1991}. 
It is now known that there are significant differences between CME and HSS-related geomagnetic activities.
E.g., CMEs are responsible for the largest geomagnetic storms \citep{Borovsky_2006} while HSSs dominate substorm activity \citep{Tanskanen_2005}.
Because of these differences, one can expect that average geomagnetic activity over suitably long time intervals can be decomposed into two components, one related to CME activity and the other related to HSS activity. 
\citet{Richardson_2000, Richardson_2002} have identified times when CMEs and HSSs were present in the solar wind at 1 AU and studied the contributions of CMEs and HSSs to the aa index.
They found that during solar maximum most aa activity is related to CMEs while during declining phase and solar minimum most aa activity is related to HSSs.
\citet{Feynman_1982} decomposed the annual aa index into two components, the `R' component being linearly related to the sunspot number and the residual `I' component defined as I = aa - R. 
While the R component is mainly produced by the CMEs the I component is more closely related to HSSs. 
This decomposition is reasonable, but it assumes, e.g., that the CME contribution to geomagnetic activity strictly follows the sunspot number, which is poorly valid around solar maxima \citep{Richardson_2012_B}.

Recently we used the principal component analysis (PCA) method to extract information on the solar wind drivers of annually averaged geomagnetic activity using a set of local $A_h$ indices \citep{Holappa_2014}. 
We found that the first principal component (PC1) represents the global average of the $A_h$ indices and correlates almost perfectly with the Ap index and that the second principal component (PC2) highly correlates with the annual fraction of high speed streams in the solar wind. 
The PCA method, however, does not decompose geomagnetic activity into pure CME and HSS components. 
For example, the first PC representing global geomagnetic activity is a mixture of CME and HSS effects, which both contribute significantly to global geomagnetic activity \citep{Richardson_2012_B}.

In this paper we develop the method further and show that the spatio-temporal information included in local indices of geomagnetic activity can be used to extract information about the independent contributions of HSSs and CMEs on geomagnetic activity without any external information about, e.g., solar activity or solar cycle phase.
We also use this information to study the contributions of the two main solar wind parameters, the solar wind speed and the interplanetary magnetic field (IMF) intensity, to geomagnetic activity. 
This paper is organized as follows. 
Section \ref{Data} introduces the $A_h$ and IHV (Inter-hourly Variability) indices used in this study.  
In Section \ref{PCA} the principal component analysis (PCA) method that we used earlier \citep{Holappa_2014} is briefly reviewed and applied to $A_h$ and IHV indices. 
The principal components are then processed using the independent component analysis (ICA) method in Section \ref{ICA}. 
The relation of the two first independent components (ICs) to solar wind speed and IMF intensity, as well as to CME and HSS fractions is discussed in Section \ref{SW_IMF}.
Finally, conclusions are given in Section \ref{concl}.

\section{Local geomagnetic indices and other data} \label{Data}

We use two different measures of local geomagnetic activity: the $A_h$ index \citep{Mursula_Martini_2007} and the IHV index \citep{Svalgaard_2007}.
The three-hourly $A_h$ index is analogous to Ak, the linearized K index \citep{Bartels_1939}, calculated from hourly data as the range of variation of the local horizontal magnetic field after removing the quiet day (Sq) variation.
However, the quiet day variation cannot be fully removed from the data by any method and some amount of residual quiet day variation also remains in the $A_h$ indices.
In order to exclude the possibility that the residual Sq variation affects our results based on the $A_h$ indices we also use $IHV$ indices which are calculated using only local night sector data and are thus practically unaffected by Sq variation. 
The daily $IHV$ index is defined as the average of six absolute hourly differences of the local horizontal magnetic field around local midnight \citep{Svalgaard_2007}.

We use the $A_h$ and the IHV indices of the 26 observatories listed in Table 1.
The selection criteria for stations was high quality and long-term continuity of their data sets and good global coverage. 
We only selected stations which have less than 20\% of data missing for any year.  
We calculated the $A_h$ and $IHV$ indices for 1966-2011 (46 years) using hourly mean data obtained from World Data Center of Edinburgh \citep{WDC-Edinburgh}.
Before calculating the indices, we checked the baselines and excluded the outliers from the magnetic data by using a three-point median filter (for more details, see \citealp{Holappa_2014}). 
We also rescaled the $A_h$ and $IHV$ indices of the CLF station for years 1966-1971 because CLF recorded spot values instead of hourly means until the end of 1971, leading to excessively large $A_h$ and $IHV$ values in these years.
For this, we calculated the averages of the ratios $A_h$(CLF)/$A_h$(NGK) and $IHV$(CLF)/$IHV$(NGK) in 1972-1981 and in 1962-1971 and multiplied $A_h$(CLF) and $IHV$(CLF) before 1971 by the corresponding ratios (0.8146 and 0.7736, respectively) so that the $A_h$(CLF)/$A_h$(NGK) and $IHV$(CLF)/$IHV$(NGK) ratios became continuous. (Note that NGK and CLF are geographically close to each other, which allows a meaningful comparison between the two stations.)

In addition to the magnetic data of ground stations, we use solar wind data from the OMNI database (\texttt{http://omniweb.gsfc.nasa.gov/}) and the classification of solar wind flow types by \citet{Richardson_2012_B}. 
There are three different solar wind types identified by \citet{Richardson_2012_B}: CMEs (including the cores of interplanetary CMEs and their related shocks and sheath regions), HSSs (corotating streams from coronal holes) and slow solar wind.

\section{Principal component analysis method}\label{PCA}

Principal component analysis \citep{Jolliffe_2005} is a statistical method, which can be used to represent a large number of correlated variables as linear combinations of a few uncorrelated variables called principal components. 
Here we apply PCA for annual means (46 years) of geomagnetic indices from 26 observatories. 
Before evaluating PCA we calculate the standardized annual means for each station separately 

\begin{equation}
A_{hs} = \frac{A_h - \langle A_h \rangle}{\sigma},
\end{equation} 

\noindent where $\langle A_h \rangle$ is the mean and $\sigma$ the standard deviation of the annually averaged $A_h$.
We calculate the standardized annual mean $IHV_s$ indices in the same way. 
Standardized annual means are then collected into the columns of the data matrix $X$ (size $46\times26$). 
PCA can be evaluated using the singular value decomposition of the data matrix (see, e.g., \citealp{Hannachi_2007})

\begin{equation}\label{svd}
X = UDV^T,
\end{equation} 

\noindent where $U$ and $V$ are orthogonal matrices ($UU^T = I$ and $VV^T = I$) and $D = \mathrm{diag}(\lambda_1, \lambda_2,...,\lambda_{26})$ contains the so called singular values of the matrix $X$. 
The column vectors of the $26\times26$ matrix $V$ are called here the empirical orthogonal functions (EOFs). 
The principal components are obtained as the column vectors of the $46\times 26$ matrix 

\begin{equation}\label{pc}
P = UD.
\end{equation} 

\noindent The original variables can then be approximated as a linear combination of the $K$ first principal components with weights given by EOFs as 

\begin{equation}\label{pca_approx}
X_{ij} = \sum_{k=1}^K P_{ik} V_{jk}
\end{equation}

\noindent where $X_{ij}$ is the value (standardized $A_h$ index) of the $j$th variable (station) at the observation time (year) $i$. 
The variance of the $k$th PC is proportional to $\lambda_k^2$. Hence, the $K$ first PCs include the following percentage

\begin{equation}\label{var_frac}
\frac{\sum_{k=1}^K \lambda_k^2}{\sum_{k=1}^{26} \lambda_k^2}\cdot 100\%
\end{equation}

\noindent of the variance in the original variables.

\subsection{The first PC}

\begin{figure}[t]
\vspace*{2mm}
\begin{center}
\includegraphics[width=12cm]{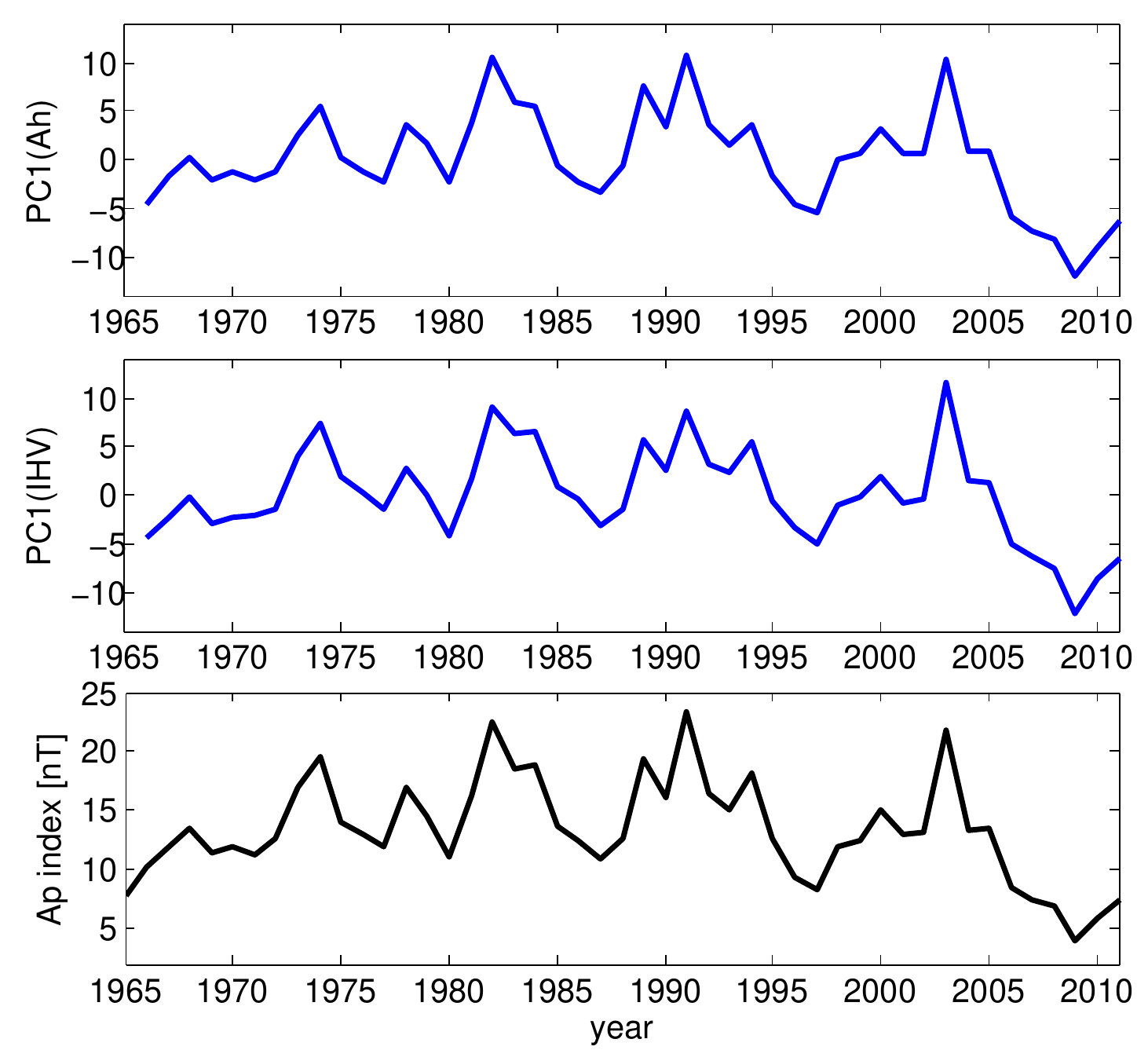}
\end{center}
\caption{The first principal component of a) $A_{hs}$ indices and b)  $IHV_s$ indices. c) The annual averages of the Ap index.}
\label{Ah_IHV_PC1}
\end{figure}

Figure \ref{Ah_IHV_PC1} shows the first principal components of the $A_{hs}$ and $IHV_s$ indices (to be called PC1(Ah) and PC1(IHV)). 
One can see that there is an excellent agreement between the PC1s of the two indices. 
The respective EOF1(Ah) and EOF1(IHV) depicted in Figure \ref{Ah_IHV_EOF} describe the latitudinal modes associated with the PC1s. 
As we found earlier \citep{Holappa_2014}, EOF1(Ah) is almost flat (independent of latitude), meaning that all stations contribute with roughly equal weights to PC1. 
Hence, the PC1(Ah) is very closely proportional to the average of the 26 $A_{hs}$ indices. 
Also the EOF1(IHV) is almost flat except for a small local minimum at the poleward boundary of the auroral oval (stations \#24 and \#25).

The PC1(Ah) and the PC1(IHV) correlate almost perfectly with the annual averages of the Ap index of the global geomagnetic activity (Pearson correlation coefficients and p-values for zero correlation from Student's t-test: cc(Ah) = 0.99, p = $6.4\cdot 10^{-34}$; cc(IHV) = 0.98, p = $2.2\cdot 10^{-31}$) which is also shown in Fig. \ref{Ah_IHV_PC1}. 
Thus, the PC1(Ah) and PC1(IHV) also closely represent the mean global geomagnetic activity.
The PC1(Ah) and PC1(IHV) already explain a large fraction of variance of the $A_{hs}$ (95.6\%) and the $IHV_s$ indices (90.1\%).  
Thus, at the annual timescale all stations at different latitudes observe roughly the same (mainly solar cycle related) long-term variation of geomagnetic activity.

\subsection{The second PC}\label{PC2}

\begin{figure}[t]
\vspace*{2mm}
\begin{center}
\includegraphics[width=12cm]{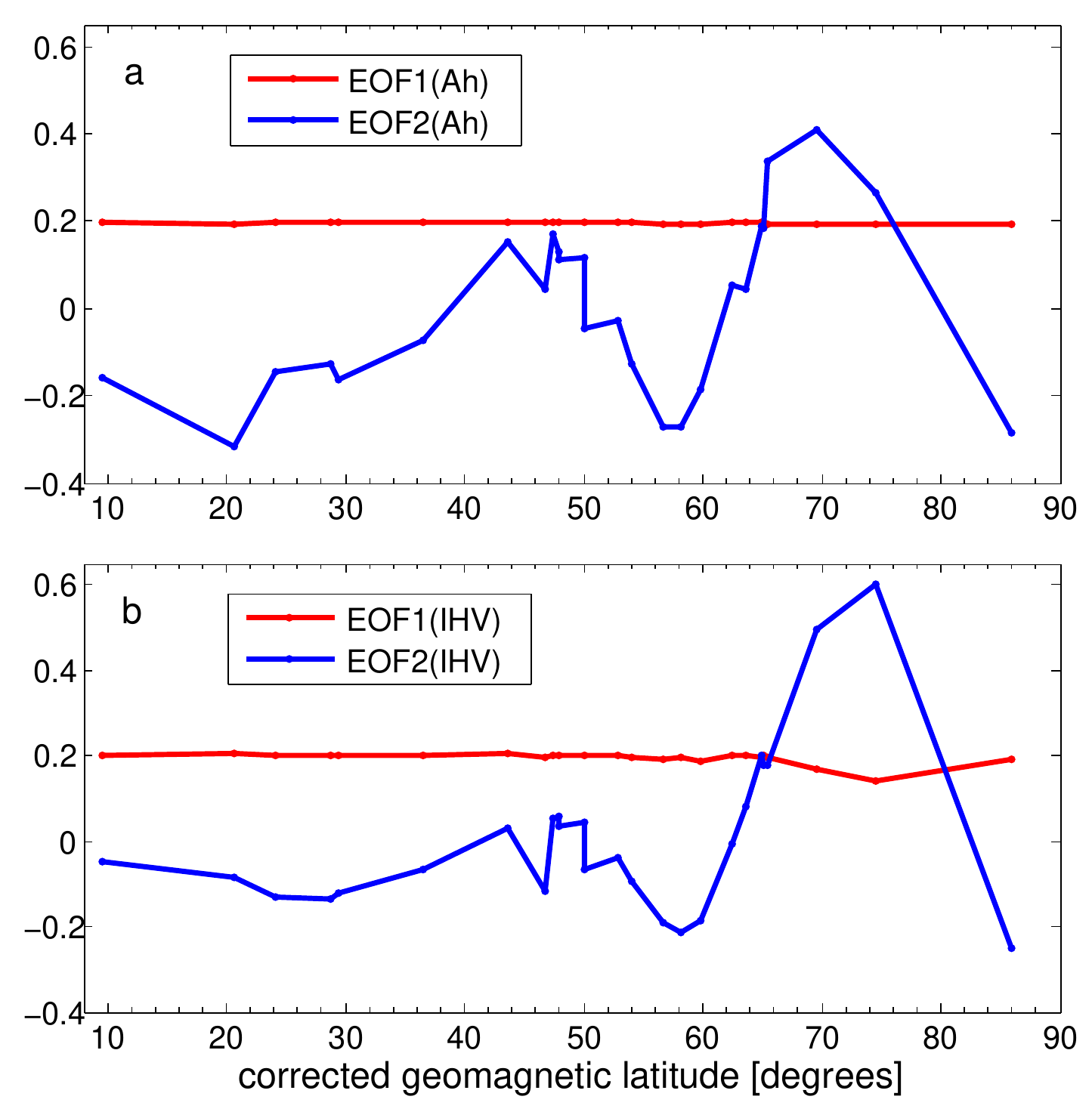}
\end{center}
\caption{Two first EOFs of a) the $A_{hs}$ and b) the $IHV_s$ indices as a function of corrected geomagnetic latitude.}
\label{Ah_IHV_EOF}
\end{figure}

\begin{figure}[t]
\vspace*{2mm}
\begin{center}
\includegraphics[width=12cm]{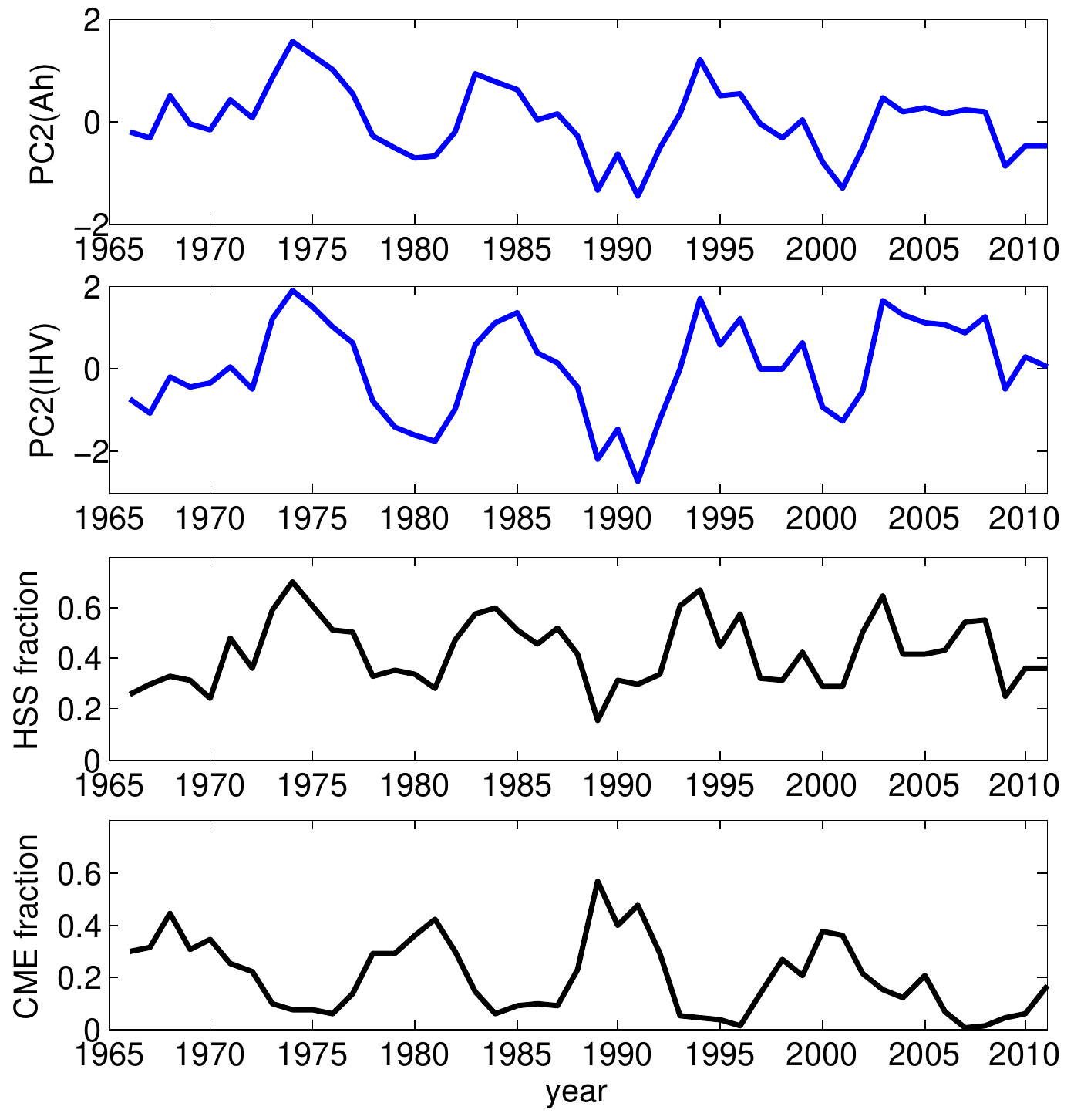}
\end{center}
\caption{a-b) The second PC of the $A_{hs}$ and the $IHV_s$ indices. c-d) Yearly fraction of HSSs and CMEs.}
\label{Ah_IHV_PC2}
\end{figure}

\begin{figure}[t]
\vspace*{2mm}
\begin{center}
\includegraphics[width=12cm]{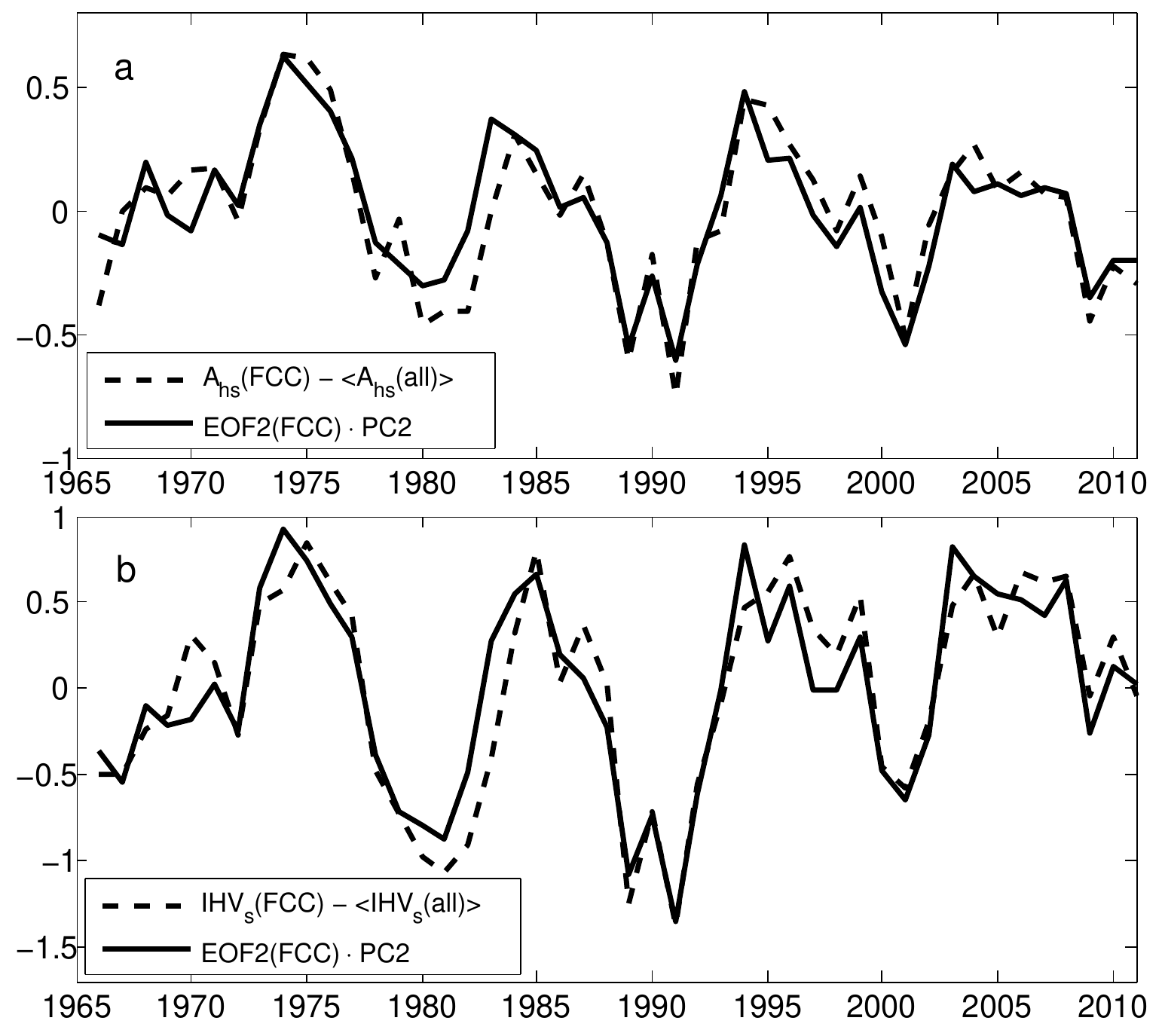}
\end{center}
\caption{a) The difference between the $A_{hs}$ index of FCC station and the global average of the $A_{hs}$ indices (solid line); and the second PC scaled by the EOF2 of FCC station (dashed line). b) The same for $IHV_s$ indices.}
\label{fcc}
\end{figure}

PC2(Ah) and PC2(IHV) are shown in Figure \ref{Ah_IHV_PC2} and the associated EOF2(Ah) and EOF2(IHV) in Figure \ref{Ah_IHV_EOF}.
As described above, the first PCs practically represent the annual global averages of the two indices. 
Therefore, the second PCs describe how these local indices at the individual stations deviate on an average from their global averages. 
For years of positive PC2(Ah) (PC2(IHV), respectively), the $A_{hs}$ ($IHV_s$) indices of stations with positive (negative) EOF2 coefficients are higher (lower) than the globally averaged $A_{hs}$ ($IHV_s$), and vice versa for years of negative PC2 values.
This is demonstrated in Figure \ref{fcc}a which shows the difference between the $A_{hs}$ index of FCC station and the average of all 26 $A_{hs}$ indices. 
For any year, $A_h$(FCC) is expected to depart from the mean of all $A_{hs}$ indices by PC2(Ah) times the EOF2 coefficient for FCC (EOF2(FCC) = 0.41).
The 2nd PC scaled by 0.41 (also shown in Fig. \ref{fcc}a) indeed explains the annual differences between the mean $A_{hs}$ and Ah(FCC) very well. 
Figure \ref{fcc}b shows an analogous difference for IHV(FCC). One can see that PC2(IHV) scaled by 0.50 (EOF2(IHV) = 0.50 for FCC) explains the annual differences between IHV(FCC) and the global $IHV_s$ very well.  

Note that the PC2 only explains 1.8\% (4.9\%) of the total variance of the $A_{hs}$ indices ($IHV_s$ indices).
Therefore, the annual deviations of individual station indices from the global average are not very large especially for stations whose EOF2 coefficients are close to zero (see Fig. \ref{Ah_IHV_EOF}). 
However, the auroral stations at $65^{\circ}-75^{\circ}$ CGM latitudes (like FCC) with the greatest positive EOF2 coefficients can notably differ from the global average. 
For example, the absolute difference between $IHV_s$(FCC) and the global mean of $IHV_s$ indices (Fig. \ref{fcc}b) can be more than one (standard deviation), which is a large difference for annual means.    

As noted earlier \citep{Holappa_2014} PC2(Ah) is very highly correlated (cc = 0.82; p = $4.6 \cdot 10^{-12}$) with the annual time fraction of high-speed streams in solar wind. 
This can also be seen in Fig. \ref{Ah_IHV_PC2} which shows the annual fraction of HSSs in solar wind according to the classification of solar wind into three flow types \citep{Richardson_2012_B}.
Figure \ref{Ah_IHV_PC2} also shows the corresponding annual fractions of CMEs which are highly anticorrelated with the HSS fractions.
Consequently, PC2(Ah) is anticorrelated with the CME fraction (cc = -0.67; p = $3.5 \cdot 10^{-7}$). 
PC2(IHV) is also very highly correlated with the HSS fraction (cc = 0.79, p = $8.2\cdot 10^{-11}$) and anticorrelated with the CME fraction (cc = -0.83; p = $9.7 \cdot 10^{-13}$).

\begin{figure}[t]
\vspace*{2mm}
\begin{center}
\includegraphics[width=12cm]{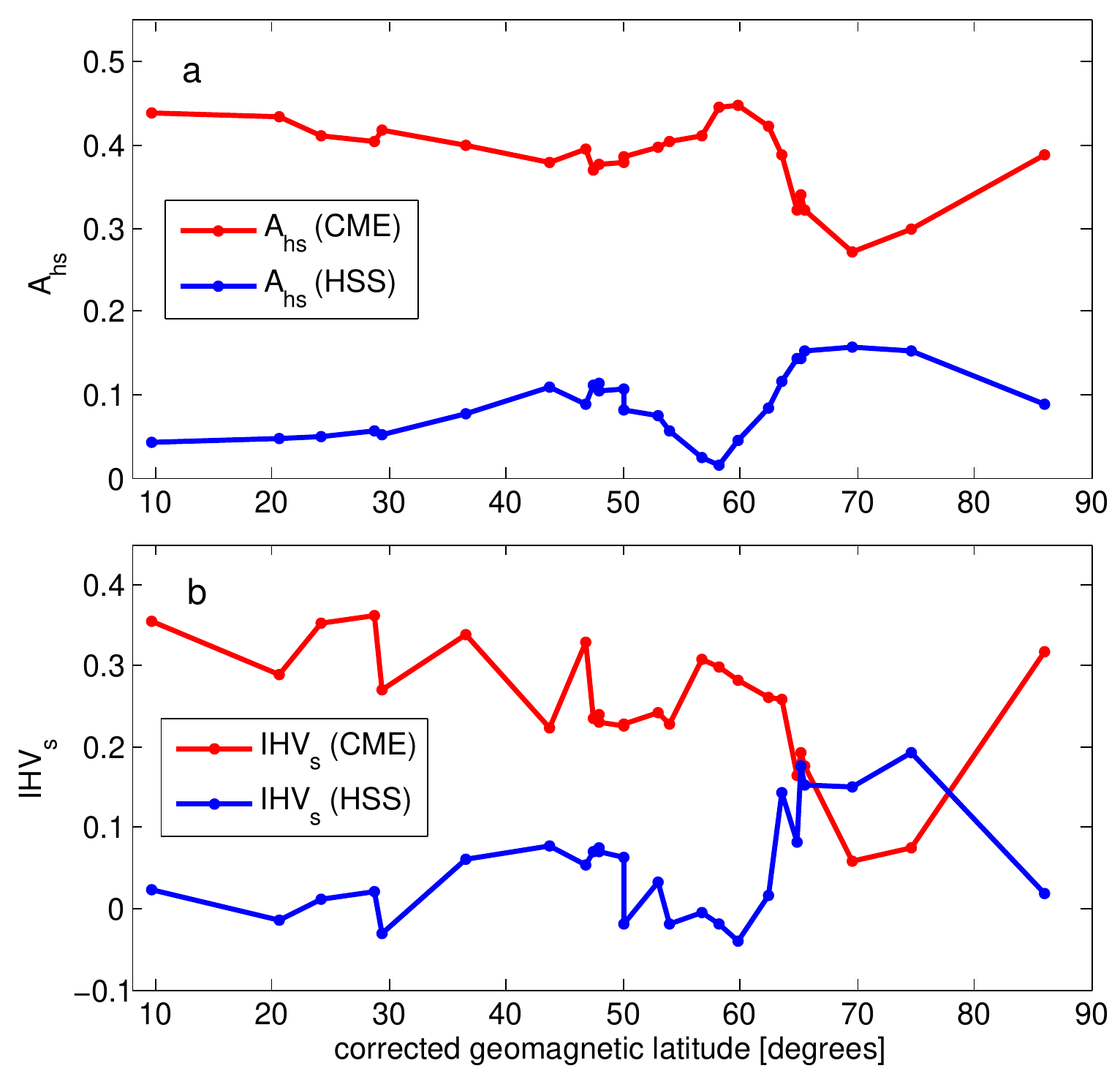}
\end{center}
\caption{Averages of the a) $A_{hs}$ and b) $IHV_s$ indices during CMEs and HSSs.}
\label{Ah_during_HSS_CME}
\end{figure}

Figure \ref{Ah_during_HSS_CME} shows the averages of the $A_{hs}$ and $IHV_s$ indices during CMEs and HSSs.  
Averages of the standardized three-hourly values of $A_{h}$ indices were calculated over those three-hour periods when only one solar wind type (CME or HSS) was present in the solar wind.
Similarly, averages of the standardized daily values of the $IHV$ indices were calculated over those local nights when only one solar wind type was present.   
As seen in Fig. \ref{Ah_during_HSS_CME}, there are clear latitudinal patterns in the $A_{hs}$ and $IHV_s$ indices during CMEs and HSSs.
One can note the high similarity between the EOF2(Ah) (see Fig. \ref{Ah_IHV_EOF}a) and the distribution of the $A_{hs}$ indices during HSSs (Fig. \ref{Ah_during_HSS_CME}a). 
The distribution of the $A_{hs}$ indices during CMEs is almost the mirror image of the HSS distribution.  
The $IHV_s$ indices during CMEs and HSSs (Fig. \ref{Ah_during_HSS_CME}b) show roughly the same patterns as the corresponding $A_{hs}$ indices. 
Also the EOF2(IHV) (see Fig. \ref{Ah_IHV_EOF}b) resembles the EOF2(Ah) (see Fig. \ref{Ah_IHV_EOF}a) and matches with the distribution of the $IHV_s$ indices during HSSs (see Fig. \ref{Ah_during_HSS_CME}b). 

Because the second PCs of the $A_{hs}$ and $IHV_s$ indices correlate (anticorrelate) with the HSS (CME) fraction and the second EOFs match with the latitudinal distributions of the indices during HSSs (CMEs), one can conclude that PC2 is (mainly) caused by the latitudinally different response of local geomagnetic activity to CMEs and HSSs.  
Figure \ref{Ah_during_HSS_CME} shows that during HSSs the strongest values of $A_{hs}$ and $IHV_s$ indices are found at the auroral latitudes ($65^{\circ}-75^{\circ}$) while during CMEs the $A_{hs}$ and $IHV_s$ indices have a (local) maximum at subauroral latitudes ($55^{\circ}-63^{\circ}$). 
We showed earlier \citep{Holappa_2014} that the relative contribution of HSS driven substorms maximizes at the auroral latitudes while the relative effect of CME driven substorms maximizes at subauroral latitudes (where substorms are observed especially during magnetic storms \citep{Tanskanen_2002, Hoffman_Gjerloev_2010}), which explains the subauroral minimum and the auroral maximum of EOF2. 
Since $IHV_s$ indices only measure geomagnetic activity at the night sector, i.e., at the preferred local time (LT) sector of substorms, they are more sensitive to substorms (and therefore to HSSs) than the $A_{hs}$ indices.  
This explains the slightly larger variation of $IHV_s$ (HSSs) between the auroral maximum and the subauroral minimum (see Fig. \ref{Ah_during_HSS_CME}b).
This also explains why EOF2(IHV) shows a higher auroral maximum than EOF2(Ah) (see Fig. \ref{Ah_IHV_EOF} and discussion later).

\section{Independent component analysis method}\label{ICA}

The basic idea of the independent component analysis (ICA) is analogous to that of the principal component analysis: initially dependent variables are presented as a linear combination of statistically independent components. 
There are numerous ways to perform ICA (see, e.g., \citeauthor{Hyvarinen_etal_2001}, 2001), but we use here the FastICA software package (\citet{Hyvarinen_1999}, \texttt{http://research.ics.aalto.fi/ica/fastica/}).

\begin{figure}[t]
\vspace*{2mm}
\begin{center}
\includegraphics[width=12cm]{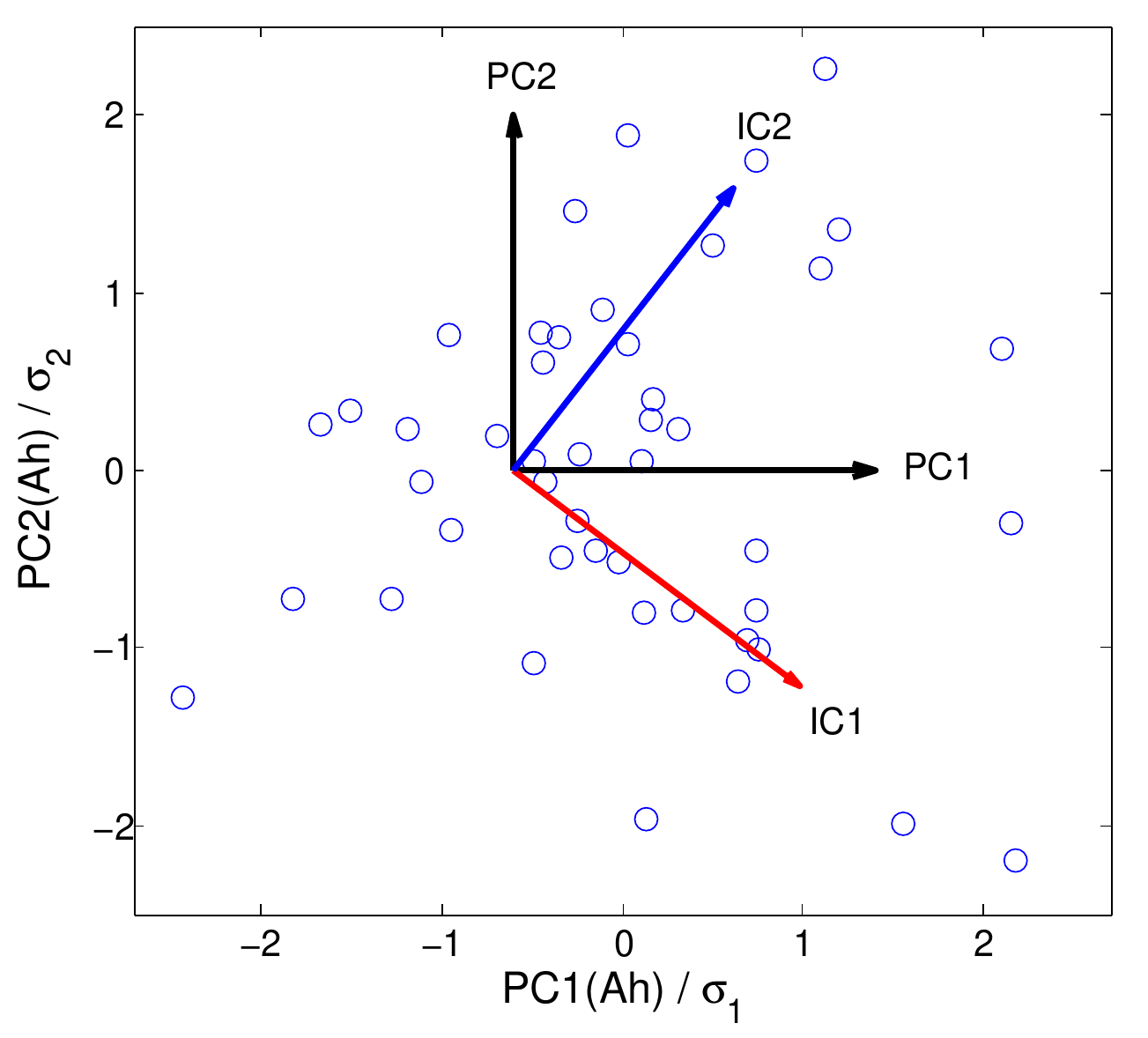}
\end{center}
\caption{Scatter plot of the standardized first and second PCs of the $A_{hs}$ indices (denoted in black). The red and blue arrows represent the row vectors of the rotation matrix $A$ on which the PCs are projected.}
\label{IC_rot}
\end{figure}

While the principal components obtained by the PCA method are uncorrelated, they are not necessarily statistically independent. 
Actually, only if the principal components are Gaussian their uncorrelatedness also guarantees their statistical independence.   
To see if the two first principal components are independent or not, we first standardize them to unit variance by dividing them by their standard deviations $\sigma_1$ and $\sigma_2$.
Using matrix notation the standardized PCs are the columns of the matrix  

\begin{equation}\label{pc_norm}
P_s = P_2 Z
\end{equation}

\noindent where the $46\times 2$ matrix $P_2$ contains the two first columns of the matrix $P$ of Eq. \ref{pc} and $Z = diag(\sigma_1^{-1}, \sigma_2^{-1})$. 
Figure \ref{IC_rot} shows a scatter plot of the standardized PC1(Ah) and PC2(Ah).
If the two PCs were statistically independent, the scatter pattern would be spherically symmetric.
Clearly this is not the case. 
One can see, e.g., that a positive value of PC1 implies either a large positive or a large negative value of PC2, and a negative value of PC1 implies a small value of PC2.
The idea of the IC analysis is to find an orthogonal rotation of the principal components that makes the rotated components statistically as independent as possible.
The rotation of the principal components can written as

\begin{equation}\label{ic}
S = A P_s^T,
\end{equation}

\noindent where the orthogonal $2\times 2$ matrix $A$ is the so called mixing matrix and the rows of $2\times 46$ matrix $S$ contain the independent components (with unit variances).
The ICA algorithm finds the matrix $A$ in an iterative process by minimizing the entropies of the independent components. 
The independent components are maximally non-Gaussian, because the Gaussian distribution has the greatest entropy among all distributions with the same variance. 

The principal components are projected onto the basis defined by the row vectors of the matrix $A$ which are shown in Figure \ref{IC_rot} as IC1 and IC2.
The matrix $A$ calculated for $A_{hs}$ indices performs a clockwise rotation by $37.4^{\circ}$, whence IC1(Ah) = $0.79 \cdot$ PC1$_{s}$(Ah) - $0.61 \cdot$ PC2$_{s}$(Ah) and IC2(Ah) = $0.61 \cdot$ PC1$_{s}$(Ah) + $0.79 \cdot$ PC2$_{s}$(Ah). 
For the $IHV_s$ indices the rotation angle is $52.3^{\circ}$, whence IC1(IHV) = $0.61\cdot$ PC1$_{s}$(IHV) - $0.79 \cdot$ PC2$_{s}$(IHV) and IC2(IHV) = $0.79 \cdot$ PC1$_{s}$(IHV) + $0.61 \cdot$ PC2$_{s}$(IHV).        

Using Equations \ref{pc_norm} and \ref{ic} the the original data matrix can be approximated as 

\begin{equation} \label{ICA_approx}
X = P_2 V^T = P_s Z^{-1} V^T = S^T A Z^{-1} V^T,
\end{equation} 

\noindent where the row vectors in the matrix $A Z^{-1} V^T$ can be interpreted as the spatial modes (SM) corresponding to the two independent components (in analogous way with the matrix $V^T$ in Eq. \ref{svd}). 
These spatial modes obtained by rotation from the EOFs in $V$, but they are not orthogonal because the matrix $A Z^{-1}$ is not orthogonal due to the different variances of the principal components. 
Equation \ref{ICA_approx} is analogous to Eq. \ref{pca_approx} and simply states that the original data can be represented as the following linear combination

\begin{equation}\label{ica_summa}
X_{ij} = \mathrm{IC}1(i) \cdot \mathrm{SM}1(j) + \mathrm{IC}2(i) \cdot \mathrm{SM}2(j),
\end{equation}

\noindent where IC1($i$) and IC2($i$) are the two independent components for year $i$ and SM1($j$) and SM2($j$) are the corresponding spatial mode coefficients for station $j$.  

ICA could also be directly applied to the original data matrix, but the ensuing ICs are not ordered according to decreasing (or increasing) importance (fraction of total variance) and thereby do not reflect the physically most important processes. Rather, in this case, ICA tends to emphasize spikes in the data, which are highly non-Gaussian, but misses the physically relevant patterns. Instead, reducing first the dimension of the data by including only the two leading PCs in the ICA makes the two ICs also to include a large fraction (95\%) of variance and the important physics.

\subsection{The first and second IC}

\begin{figure}[t]
\vspace*{2mm}
\begin{center}
\includegraphics[width=12cm]{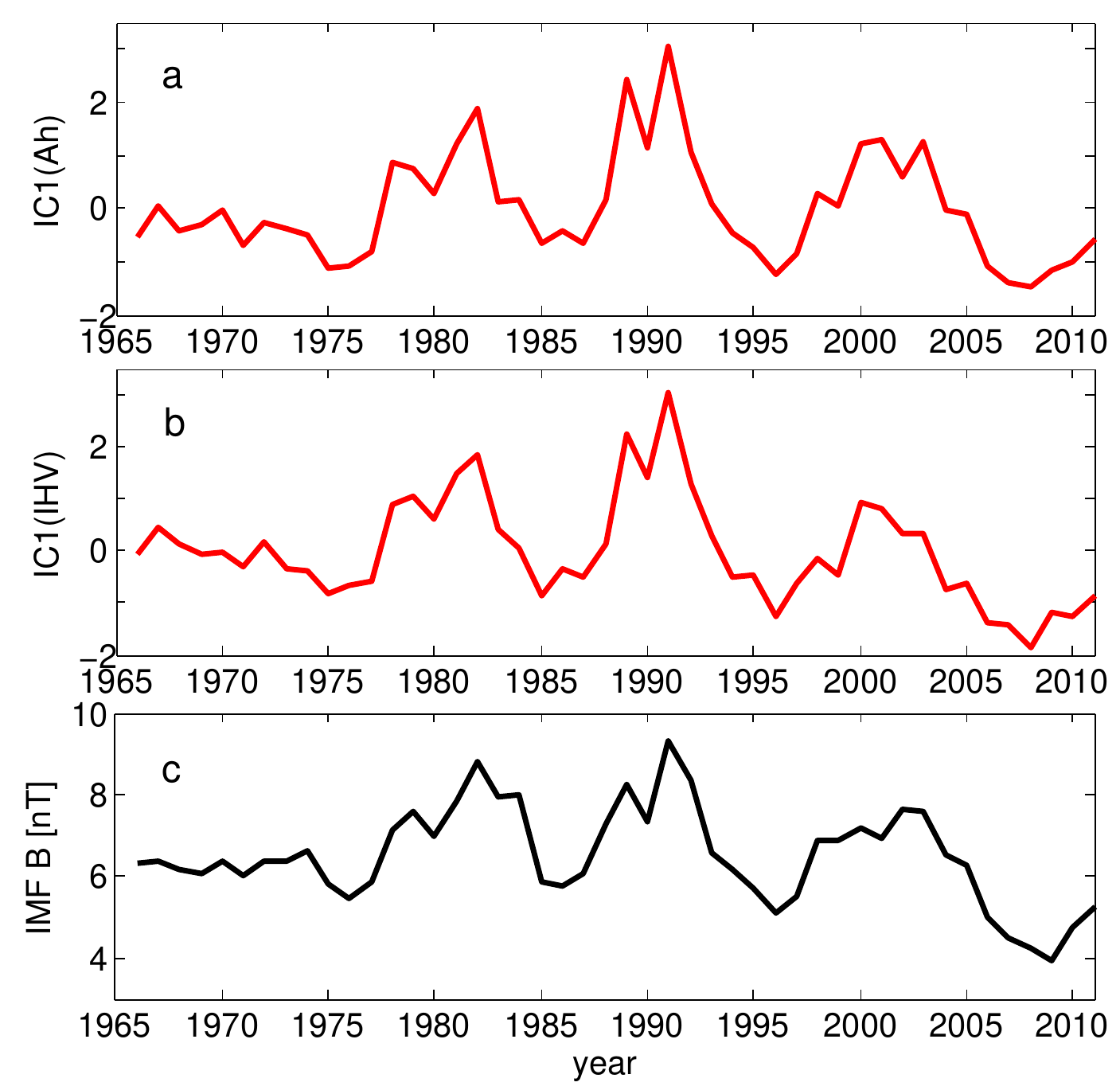}
\end{center}
\caption{a-b) The first ICs of the $A_{hs}$ and the $IHV_s$ indices. c) The annual averages of the IMF strength $B$.}
\label{Ah_IHV_IC1}
\end{figure}

\begin{figure}[t]
\vspace*{2mm}
\begin{center}
\includegraphics[width=12cm]{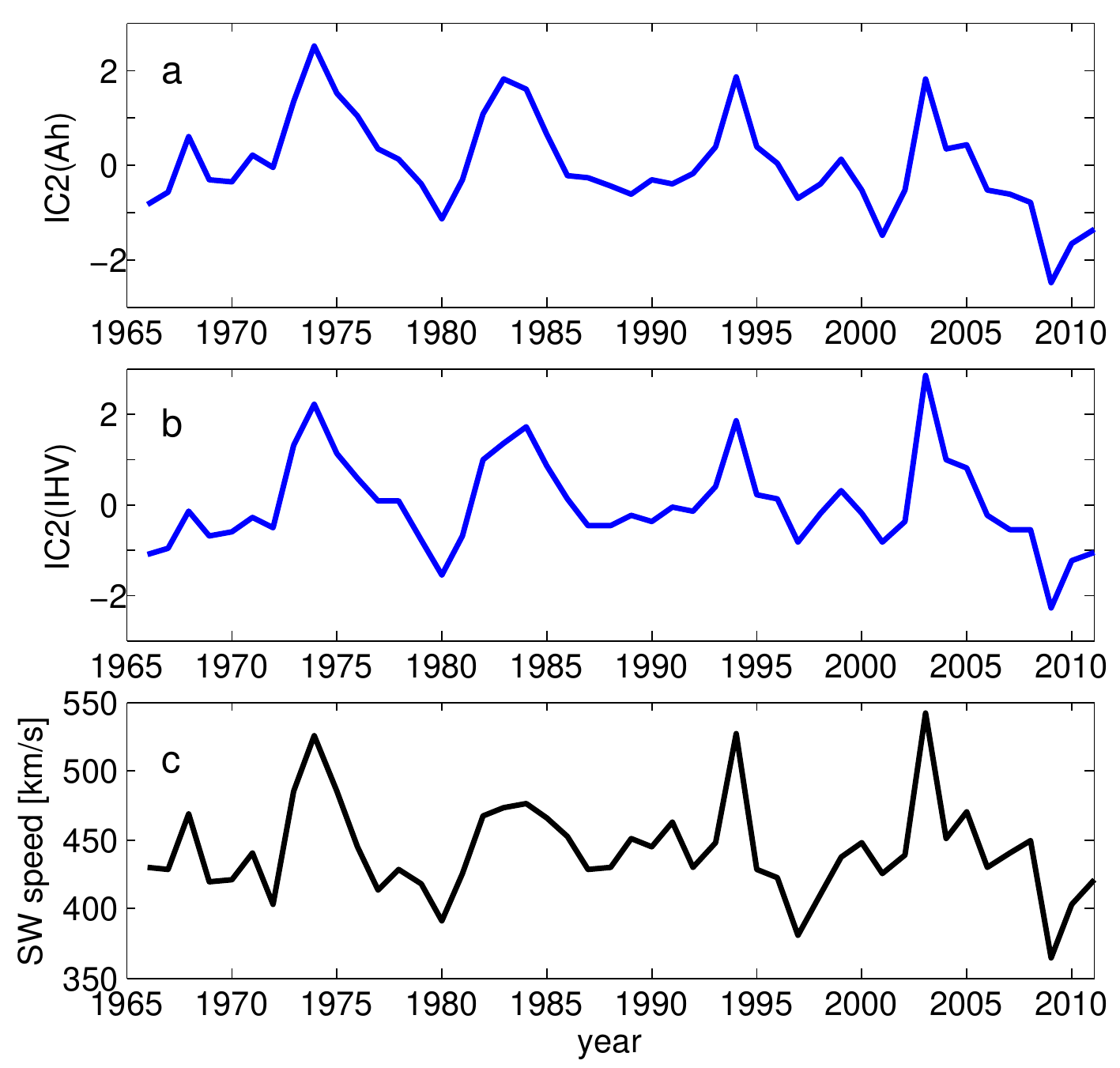}
\end{center}
\caption{a-b) The second ICs of the $A_{hs}$ and the $IHV_s$ indices. c) The annual averages of the solar wind speed.}
\label{Ah_IHV_IC2}
\end{figure}

Figures \ref{Ah_IHV_IC1} and \ref{Ah_IHV_IC2} show the first and second independent components for the two indices, respectively.
One can see that the ICs of the two indices are very similar with each other, as expected from the similarity of the two first PCs of these indices.
The correlations between the ICs of the two indices are very high: cc(IC1(Ah), IC1(IHV)) = 0.95, p = $4.9\cdot 10^{-23}$ and cc(IC2(Ah), IC2(IHV)) = 0.94, p = $1.1\cdot 10^{-21}$. 

The spatial modes corresponding to the two ICs are depicted in Figure \ref{Ah_IHV_IC_eof}.
One can see that the two spatial modes are almost mirror images of each other for both indices, especially for $A_{hs}$. 
However, the first spatial mode of IHV shows a very deep minimum at auroral latitudes, which is also related to the dip in EOF1(IHV) (Fig. \ref{Ah_IHV_EOF}).
Note also that the SM2(IHV) is generally larger than SM1(IHV). 
This means that the second IC has, on the average, a higher weight in the IHV indices than the first IC. 
This is opposite to $A_h$ indices for which the first IC is dominating.

\section{Relation to solar wind and IMF} \label{SW_IMF}

Annual averages of the IMF intensity $B$ and the solar wind speed $v$ are plotted in Figures \ref{Ah_IHV_IC1}c and \ref{Ah_IHV_IC2}c, respectively.
One can see that IC1(Ah) and IC1(IHV) are very highly correlated with the IMF intensity $B$ with cc(IC1(Ah), $B$) = 0.90; p = $4.2\cdot 10^{-17}$ and cc(IC1(IHV), $B$) = 0.85; p = $1.8\cdot 10^{-12}$.
The second ICs are, in turn, very highly correlated with the solar wind speed $v$: cc(IC2(Ah), $v$) = 0.82; p = $4.7\cdot 10^{-13}$ and cc(IC2(IHV), $v$) = 0.89; p = $5.0\cdot 10^{-16}$, 
or alternatively with $v^2$: cc(IC2(Ah), $v^2$) = 0.81; p = $6.2\cdot 10^{-12}$ and cc(IC2(IHV), $v^2$) = 0.89; p = $2.8\cdot 10^{-16}$.
These correlations and the above ICA results expressed in Eq. (\ref{ica_summa}) suggest that the annual averages of all local geomagnetic indices can be represented as a linear combination of the annual solar wind speed and the IMF strength with their own optimum relative weights for these two drivers.

Before presenting the results we note that, of course, it is not physically reasonable that momentary geomagnetic activity should depend on a linear combination of $B$ and $v$ (or $v^2$). 
Rather, the relation between geomagnetic activity and solar wind parameters is usually expressed in terms of different nonlinear coupling functions, e.g., $Bv^2$. 
There are also many coupling functions involving, e.g., solar wind density and IMF vector orientation, but at the annual timescale they do not correlate any better with global geomagnetic activity than the simple function $Bv^2$ \citep{Finch_2007}.
The above ICA results and the earlier results regarding the nonlinear solar wind coupling functions can be understood as follows. 
During CMEs the coupling function $Bv^2$ is mainly enhanced above the mean value due to large values of $B$, with $v$ remaining at the average level, while during HSSs the high values of $Bv^2$ are due to persistently high values of $v$, with $B$ attaining average values \citep{Richardson_2002, Richardson_2012_B}.

To further test this hypothesis, we decompose hourly $B$ and $v^2$ values into constant and fluctuating parts: $B = B_0 + B'$ and $v^2 = v^2_0 + (v^2)'$, where $B_0$ and $v_0^2$ denote the averages of $B$ and $v^2$ in 1966-2011.  
Now we can write 


\begin{equation}\label{v2b_decomp}
Bv^2  =  B_0 v_0^2 + B' v_0^2 + B_0 (v^2)' + B' (v^2)'.
\end{equation}  

\begin{figure}[t]
\vspace*{2mm}
\begin{center}
\includegraphics[width=12cm]{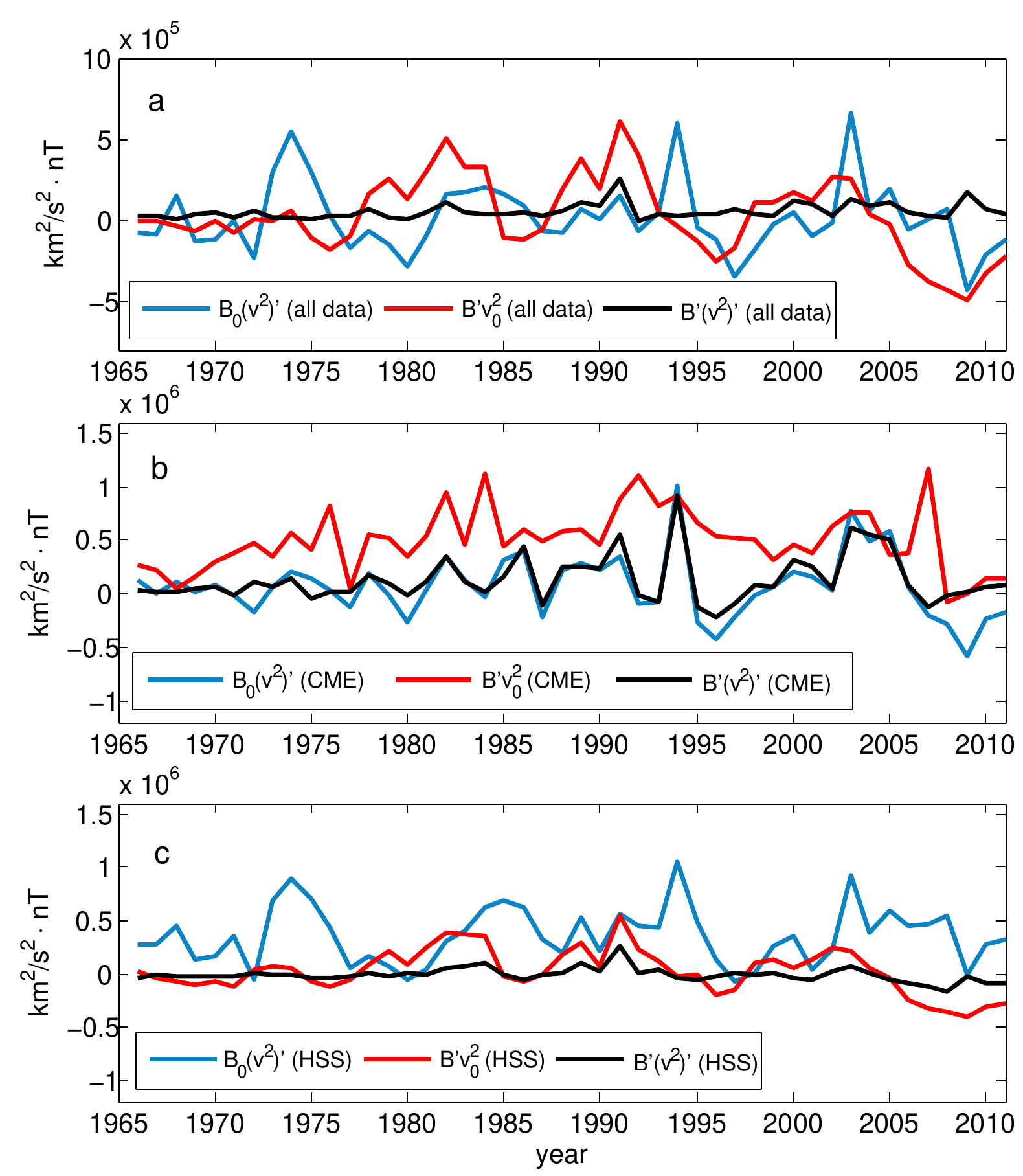}
\end{center}
\caption{Annual averages of the three time dependent terms in Equation \ref{v2b_decomp} contributing to the coupling function $Bv^2$ during a) all times b) CME c) HSS intervals.}
\label{v2b_fluct}
\end{figure}

\noindent The first term on the right hand side determines the average value of the coupling function over the 46 year period ($B_0 v_0^2 = 1.3\cdot10^{6}$ nT$\cdot$km$^2$/$s^2$, $B_0 = 6.4$ nT, $v_0$ = 439km/s), which, however, does not affect, e.g., the correlation between the coupling function and geomagnetic activity.
Figure \ref{v2b_fluct}a shows the annual averages of the three last time-dependent terms on the right hand side of Eq. \ref{v2b_decomp} including all solar wind data.
One can see that the third term $B' (v^2)'$ is overall rather small, suggesting that the fluctuations $B'$ and $(v^2)'$ (and in fact also $B$ and $v^2$) are rather uncorrelated.  
This also leads to the fact that, at the annual time scale, the functional form of the coupling function $Bv^2$ can indeed be effectively represented as a linear combination of $B$ and $v^2$.
Hence, at the annual timescale, geomagnetic activity has two components, one correlated with the IMF strength and the other with the solar wind speed.
Both fluctuating terms in Fig. \ref{v2b_fluct}a have approximately the same range of variation meaning that $B$ and $v^2$ contribute to the variations of the coupling function $Bv^2$ roughly equally.  
  
Figures \ref{v2b_fluct}b and \ref{v2b_fluct}c show the annual averages of the three time-dependent terms of Eq. \ref{v2b_decomp} during CMEs and HSSs, respectively.
One can see that the term $B'v_0^2$ clearly dominates over the two other terms during CMEs while the term $B_0(v^2)'$ dominates during HSSs.
Therefore, the IMF strength is indeed the dominant parameter driving global geomagnetic activity during CMEs, while the solar wind speed dominates during HSSs.
Interestingly, in 1994 and 2003 all three terms are high during CMEs indicating that in these years CMEs carried strong magnetic fields and were very fast.

\subsection{Relation to CMEs and HSSs} \label{CME_HSS}

\begin{figure}[t]
\vspace*{2mm}
\begin{center}
\includegraphics[width=12cm]{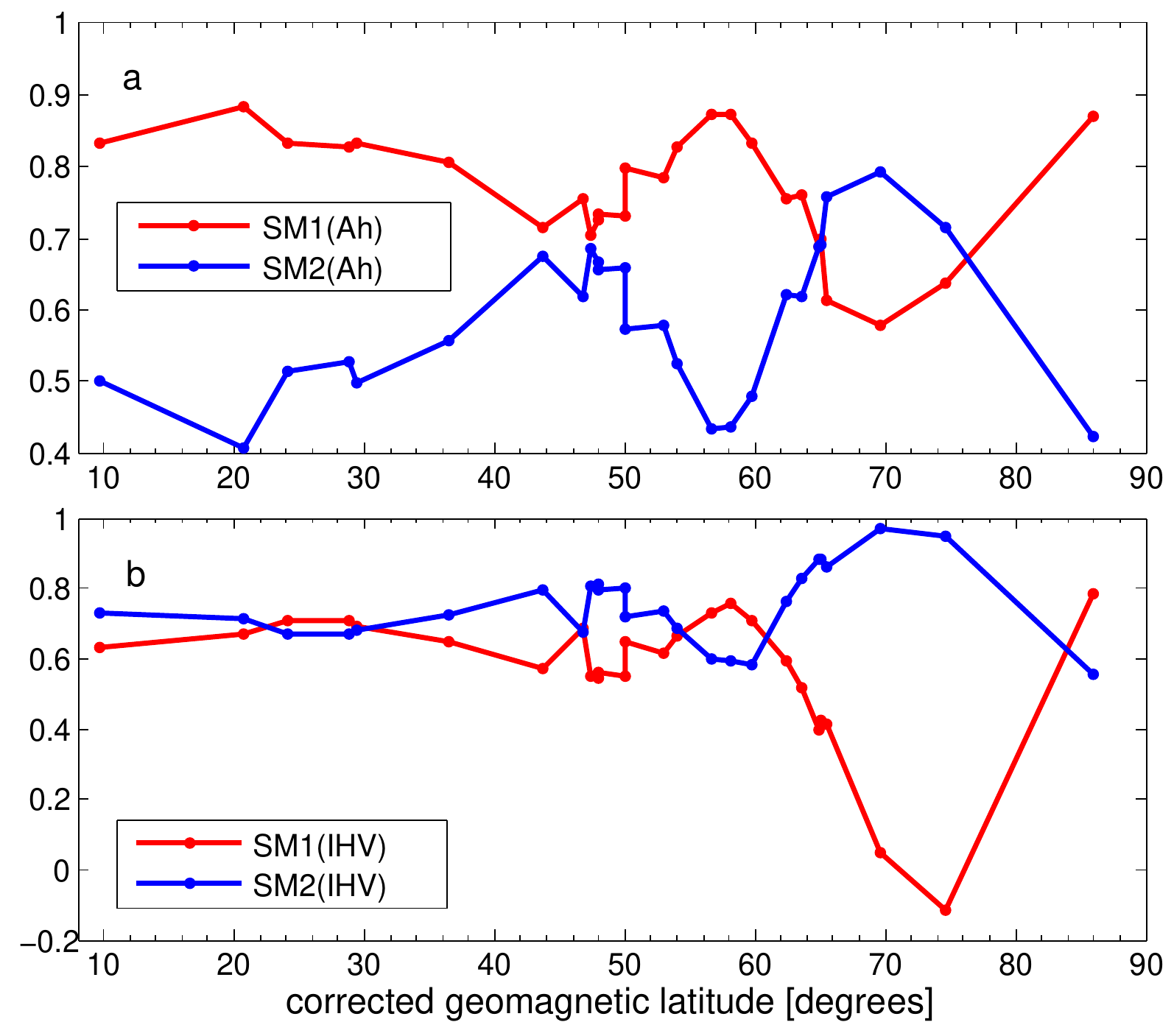}
\end{center}
\caption{The spatial modes corresponding to the two ICs of a) $A_{hs}$ and b) $IHV_s$ indices as functions of corrected geomagnetic latitude.}
\label{Ah_IHV_IC_eof}
\end{figure}

\begin{figure}[t]
\vspace*{2mm}
\begin{center}
\includegraphics[width=12cm]{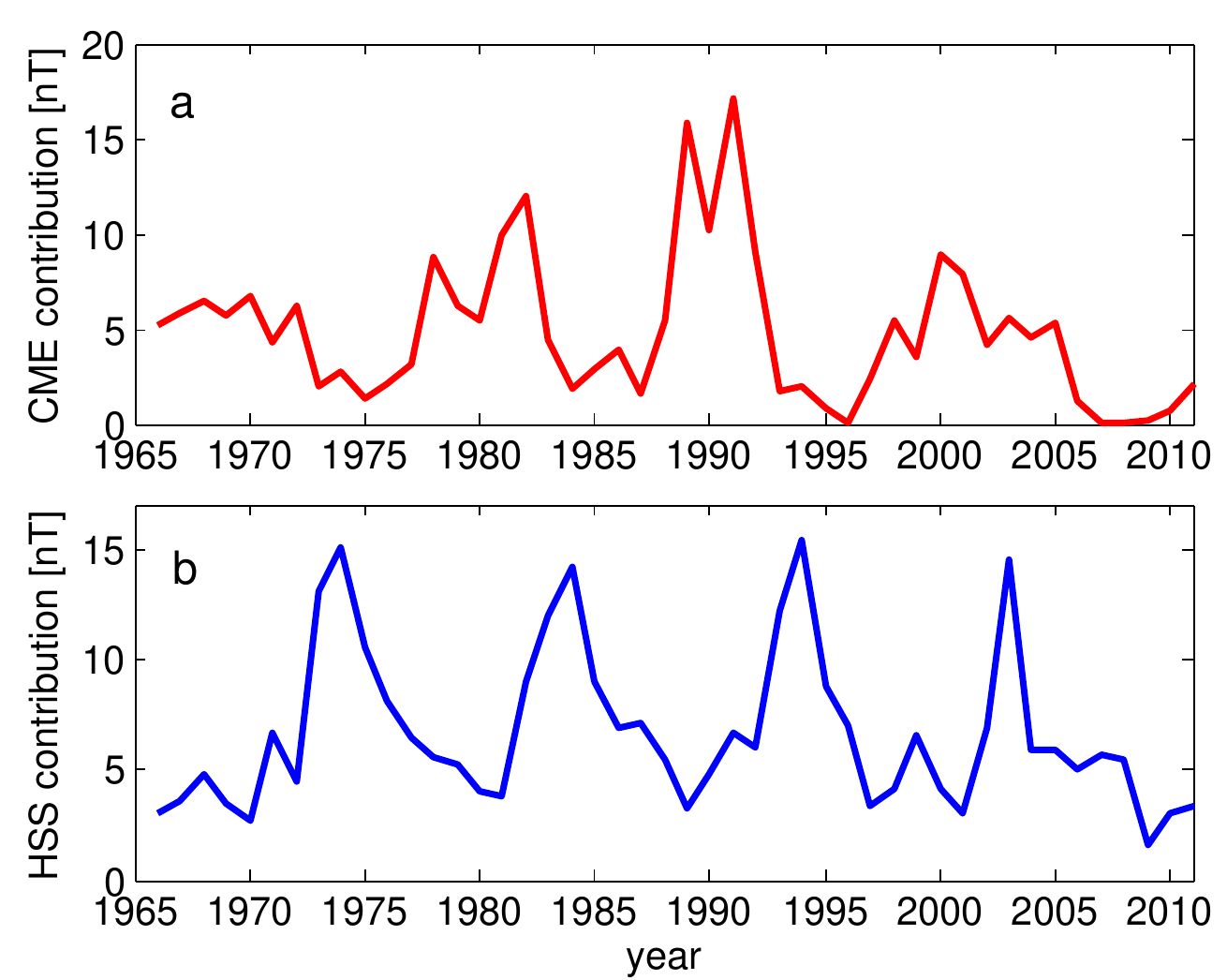}
\end{center}
\caption{Annual a) CME and b) HSS contributions to the Ap index.}
\label{CME_HSS_contrib}
\end{figure}

The ICA spatial modes in Figure \ref{Ah_IHV_IC_eof} have a quite similar latitudinal patterns as the average distributions of the $A_{hs}$ and $IHV_s$ indices during CMEs and HSSs depicted in Figure \ref{Ah_during_HSS_CME}. 
This suggests that the IC1(Ah) and IC1(IHV) represent the CME contributions to these indices while the second ICs represent the HSS contributions. 
The first ICs correlate well with the CME fraction (cc(IC1(Ah)) = 0.76, p = $6.3 \cdot 10^{-10}$; cc(IC1(IHV)) = 0.81, p = $5.6 \cdot 10^{-12}$) and the second ICs with the HSS fraction (cc(IC2(Ah)) = 0.73, p = $7.8 \cdot 10^{-9}$; cc(IC2(IHV)) = 0.74, p = $4.5 \cdot 10^{-9}$).
However, it is not physical that the annual fractions of CMEs and HSSs in solar wind should determine the yearly levels of geomagnetic activity because the properties of CMEs and HSSs evolve from one year to another.
For example, as shown clearly in Fig. \ref{v2b_fluct}, the speeds and magnetic field strengths of CMEs and HSSs are different in different years.   
To take the varying properties of CMEs and HSSs into account, we estimate the CME and HSS contributions to global geomagnetic activity by calculating the quantities 

\begin{eqnarray}
C &=& \langle Ap \rangle_{CME} \cdot f_{CME}\\
H &=& \langle Ap \rangle_{HSS} \cdot f_{HSS},
\end{eqnarray}

\noindent where $\langle Ap \rangle_{CME}$ ($\langle Ap \rangle_{HSS}$) is the annual average of the Ap index values observed during CMEs (HSSs) and $f_{CME}$ ($f_{HSS}$) is the annual fraction of CMEs (HSSs) in the solar wind, and plotting them in Figure \ref{CME_HSS_contrib}.  
As expected, the first ICs (see Fig. \ref{Ah_IHV_IC1}) are very highly correlated with the CME contribution (cc(IC1(Ah)) = 0.92, p = $5.7\cdot 10^{-19}$; cc(IC1(IHV)) = 0.93, p = $3.3\cdot 10^{-20}$) and the second ICs (see Fig. \ref{Ah_IHV_IC2}) with the HSS contribution (cc(IC2(Ah)) = 0.88, p = $4.7\cdot 10^{-16}$; cc(IC2(IHV)) = 0.90, p = $4.9\cdot 10^{-17}$). 
This gives strong evidence that the first and second ICs indeed represent the contribution of CMEs and HSSs, respectively, to geomagnetic activity.
There are some small differences, e.g., between the second ICs and the HSS contribution, especially in 1989, when the HSS contribution shows a deep minimum but the second ICs only a shallow minimum. 
These differences are most likely related to the numerous gaps in the solar wind satellite measurements in 1980s and early 1990s, causing larger inaccuracy in solar wind classification and in the annual CME and HSS fractions at those times \citep{Richardson_2012_B}.   

Since the two ICs represent the CME and HSS contributions to geomagnetic activity, the corresponding IC spatial modes quantify the weights by which CMEs and HSSs contribute to the local geomagnetic activity at the different stations. 
Although the spatial modes of $A_{hs}$  and $IHV_s$ indices (see Fig. \ref{Ah_IHV_IC_eof}) have a fairly similar latitudinal variation, the SM2(IHV) is at considerably higher level than SM2(Ah), indicating that the relative contribution of HSSs is, on an average, greater to the $IHV_s$ indices than to the $A_{hs}$ indices. 
Furthermore, the first spatial mode of $IHV_s$ shows a very deep minimum at the poleward edge of the auroral oval, meaning that CMEs have a very small contribution to the $IHV_s$ indices at these latitudes where geomagnetic activity is dominated by HSS-driven substorm activity in the night sector \citep{Tanskanen_2005, Tanskanen_2011}.
This is also consistent with the results by \citet{Finch_2008} who showed that correlation between geomagnetic activity and solar wind speed maximizes in the night sector at auroral latitudes.
On the other hand, the $A_{hs}$ indices measure all local times and are thus not solely dominated by substorms even at auroral latitudes, which decreases the relative importance of HSSs in the $A_{hs}$ indices.   
Because of the strong dominance of HSSs, the $IHV_s$ indices at auroral latitudes have a slightly higher EOF2 and a smaller EOF1 (see Fig. \ref{Ah_IHV_EOF}), as discussed in Section \ref{PC2}.

\begin{figure}[t]
\vspace*{2mm}
\begin{center}
\includegraphics[width=12cm]{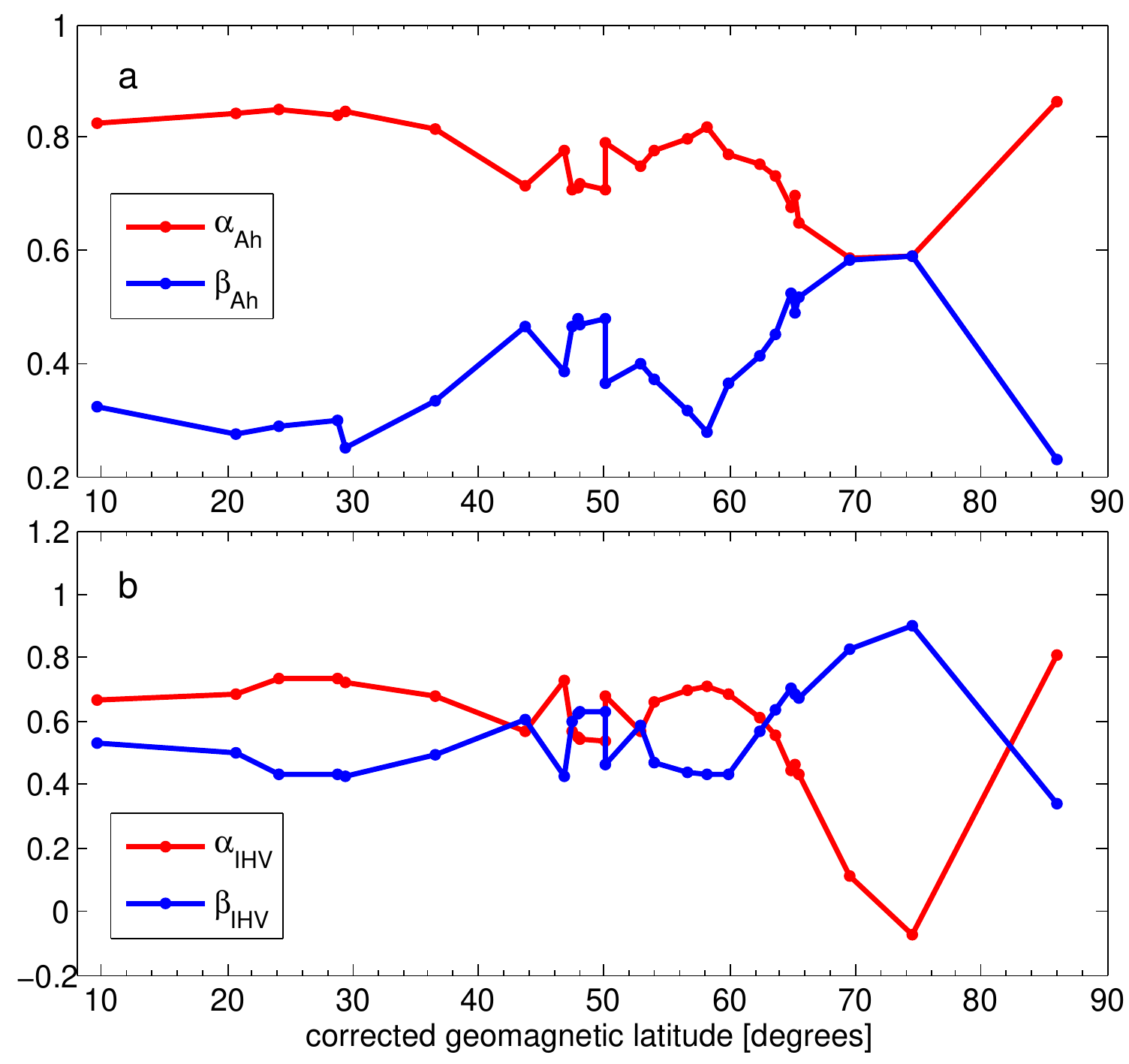}
\end{center}
\caption{Least squares fit coefficients $\alpha$ and $\beta$ (Eqs. \ref{fit}-\ref{fit_ihv}) for a) the $A_{hs}$ indices and b) the $IHV_s$ indices.}
\label{fit_coeff}
\end{figure}

In order to exclude the possibility that the spatial modes obtained by the independent component analysis are artifacts of the method, we have fitted coefficients $\alpha$ and $\beta$ for $A_{hs}$ and $IHV_s$ indices of different stations so that  

\begin{eqnarray} \label{fit}
A_{hs} &=& \alpha_{Ah} B_s + \beta_{Ah} v_s^2\\ 
IHV_s &=& \alpha_{IHV} B_s + \beta_{IHV} v_s^2, \label{fit_ihv}
\end{eqnarray} 

\noindent where $B_s$ and $v_s^2$ are standardized IMF strength and squared solar wind speed, respectively. 
The coefficients $\alpha_{Ah} (\alpha_{IHV}$) and $\beta_{Ah} (\beta_{IHV}$) are solved using the standard least squares fitting method. As seen in Fig. \ref{fit_coeff}, coefficients of Eq. \ref{fit} have the same latitudinal variation as the ICA spatial modes (Eq. \ref{ICA_approx}). 
Thus, the coefficients $\alpha_{Ah}$ ($\alpha_{IHV}$) and $\beta_{Ah}$ ($\beta_{IHV}$) obtained from the least squares fits are very similar with the first and second spatial mode coefficients of $A_{hs}$ ($IHV_s$) indices, respectively.
The only systematic difference is that the $\beta_{Ah}$ coefficients are somewhat smaller than the coefficients of SM2(Ah). 
The fact that the least squares fit calculated using the measured solar wind data produces very similar results with the ICA (blind to solar wind data) gives great confidence on the results based on ICA method and their interpretation.

\section{Conclusions} \label{concl}  

In this paper we have studied the spatio-temporal evolution of geomagnetic activity in 1966-2011 using local $A_h$ and $IHV$ indices of 26 stations covering a wide range of latitudes. 
We analyzed the indices using the principal component analysis method and confirmed that our recent results for the $A_h$ indices \citep{Holappa_2014} also hold for $IHV$ indices, i.e., that the first PC describes global average geomagnetic activity and the second PC the deviations from the global average caused by high speed streams.

We used the independent component analysis method to rotate the two first PCs into two independent components (ICs).
The spatial modes of the two ICs clearly correspond to the distribution of the indices during CMEs (first mode) and HSSs (second mode).
The two first ICs were found to match very well with the CME and HSS contributions to global geomagnetic activity.
We also found that the first IC and the second IC correlate very highly with the IMF strength and the solar wind speed, respectively.    
This is due to the fact that high values of the IMF strength mainly dominate the (larger than average) driving of geomagnetic activity during CMEs while high solar wind speed dominates the driving during HSSs.

We found essentially similar results both for $A_h$, which include all local times and for $IHV$ indices, which only include the night sector. 
This shows that the residual Sq variation in the $A_h$ indices has no major effect to the main results. 
It is also very reassuring that the same results can be found using indices which define geomagnetic activity quite differently: the $A_h$ being a traditional range index, the $IHV$ index using hourly absolute differences.
Despite all these differences between the two indices, the PC and IC methods are able to find essentially the same information about the solar wind drivers. 
The combined PC/IC method presented here offers a new way to gain information about the relative occurrence of CMEs and HSSs and the long-term properties of solar wind, in particular the IMF strength and the solar wind speed.
This improves our understanding of the long-term evolution of solar wind and the long-term driving of geomagnetic activity by the different solar wind structures.

\begin{acknowledgments}
We acknowledge the financial support by the Academy of Finland to the ReSoLVE Centre of Excellence (project no. 272157) and to project no. 257403.
This work has benefited for collaborations and contacts within the COST ES1005 (TOSCA) Network Action (esp. Working Group 2).
The hourly magnetometer data were obtained from the World Data Center for Geomagnetism, Edinburgh (\texttt{http://www.wdc.bgs.ac.uk/}).
The Ap index was obtained from World Data Center for Geomagnetism, Kyoto (\texttt{http://http://wdc.kugi.kyoto-u.ac.jp/}).
The hourly values of the solar wind speed were obtained from the OMNI database (\texttt{http://omniweb.gsfc.nasa.gov/}).
The list of the solar wind structures can be obtained by contacting Ian G. Richardson.

\end{acknowledgments}

\end{article}



%
%
%
%
%
%

\newpage


\begin{table}
\begin{center}
\begin{tabular}{ll|cccc}
\# & Station name and code & GG lat & GG long & CGM lat & CGM long \\\hline
1&Alibag (ABG) & 18.638  & 72.872 & 9.52 & 145.27 \\
2&MBour (MBO) & 14.384 &  -16.967 & 20.78 & 56.717 \\
3&Kanoya (KNY) & 31.420 & 130.882 & 24.17 & 202.020 \\
4&Kakioka (KAK) & 36.233 &  140.183 & 28.78 & 210.93 \\
5&San Juan (SJG) & 18.382 &  -66.118 & 29.27 & 5.02 \\
6&Memambetsu (MMB) & 43.907 & 144.193 & 36.56 & 214.56 \\
7&Chambon-la-Foret (CLF) & 48.017 & 2.267 & 43.67 & 79.94 \\
8&Irkutsk (IRT) & 52.167 & 104.450 & 46.78 & 176.67 \\
9&Belsk (BEL) & 51.837 & 20.792 & 47.41 & 96.38 \\
10&Niemegk (NGK) & 52.072 & 12.675 & 47.93 & 89.65 \\
11&Hartland (HAD) & 51.000 &  -4.483 & 47.99 & 75.55 \\
12&Wingst (WNG) & 53.743  &  9.073  & 50.05 & 87.31 \\
13&Fredericksburg (FRD) & 38.210 & -77.367 & 50.07 & 356.16 \\
14&Eskdalemuir (ESK) & 55.317 & -3.200 & 52.95 & 78.22 \\
15&Victoria (VIC) & 48.517 & -123.417 & 54.04 & 294.56 \\
16&Nurmij\"arvi (NUR) & 60.508 &  24.655 & 56.69 & 102.78 \\
17&Lerwick (LER) & 60.133 & -1.183 & 58.16 & 82.11 \\
18&Sitka (SIT) & 57.052 & -135.335 & 59.82 & 278.10 \\
19&Meanook (MEA) & 54.615 & -113.347 & 62.41 & 303.72 \\
20&Sodankyl\"a (SOD) & 67.367 &  26.633 & 63.64 & 108.17 \\
21&College (CMO) & 64.867 & -147.860 & 64.88 & 261.68 \\
22&Abisko (ABK) & 68.358  & 18.823 & 65.11 & 102.91 \\
23&Leirvogur (LRV) & 64.183 &  -21.7 & 65.46 & 68.57 \\
24&Fort Churchill (FCC) & 58.786 & -94.088 & 69.61 & 330.03 \\
25&Baker Lake (BLC) & 64.333 & -96.033 & 74.59 & 324.68 \\
26&Thule (THL) & 77.483 & -69.167 & 86.00 & 36.77 \\
\end{tabular}\caption{Stations and their geographic (GG) and corrected geomagnetic (CGM) latitudes and longitudes. Stations are ordered according to their CGM latitudes.}
\label{stationtable}
\end{center}
\end{table}

\newpage


\end{document}